# Properties of Stochastic Kronecker Graph

**Ahmed Mehedi Nizam[1], Md. Nasim Adnan[2], Md. Rashedul Islam[2] and Mohammad Akbar Kabir[3]**

**[1]Detp. of CSE, Bangladesh University of Engineering and Technology**
**Dhaka 1000, Bangladesh**
*ahmed_mehedi_nizam@yahoo.com*

**[2]Dept. of CSE, University of Liberal Arts Bangladesh**
**Dhaka 1209, Bangladesh**
*nasim.adnan@ulab.edu.bd, rashedul.islam@ulab.edu.bd*

**[3]Dept. of Economics, University of Dhaka**
**Dhaka, Bangladesh**
*akbar_kabir03@yahoo.com*

**Abstract**

The stochastic Kronecker Graph model can generate large random graph that closely resembles many real world networks. For example, the output graph has a heavy-tailed degree distribution, has a (low) diameter that effectively remains constant over time and obeys the so-called densification power law [1]. Aside from this list of very important graph properties, one may ask for some additional information about the output graph: What will be the expected number of isolated vertices? How many edges, self loops are there in the graph? What will be the expected number of triangles in a random realization? Here we try to answer the above questions. In the first phase, we bound the expected values of the aforementioned features from above. Next we establish the sufficient conditions to generate stochastic Kronecker graph with a wide range of interesting properties. Finally we show two phase transitions for the appearance of edges and self loops in stochastic Kronecker graph.

## 1. Introduction

Suppose we have designed a communication protocol for the internet and are very eager to know how well it will perform in the next five years. What we need to do is to simulate our protocol on the graph that is unknown today and yet changing in the most unpredictable manner over the time. Fortunately enough, it has been observed that network in the real world does not grow/shrink in a truly *random* manner. Rather this evolves in such a way as to give birth to a graph that has a power law degree distribution (power laws have been found in the internet [2], the Web [3], citation graph [4], online social network [5] and in many others), small effective diameter (effective diameter has been found to be small for massive real world networks like the Internet, the Web and Online Social Network [6] ) and so on. So we need a suitable graph generator that can generate graphs with power law degree distribution and small effective diameter. Such a graph generator has recently been proposed in [1] and it is based on a non-standard matrix operation, namely: the Kronecker product. The model starts with an initiator graph $G_1$ having $N_1$ nodes and $E_1$ edges and (gradually) computes the $k^{th}$ Kronecker product $G_k$ of it. The graph $G_k$ would have $N_1^k$ nodes and $E_1^k$ edges and thus exhibits a version of Densification Power Law. Additionally it will be a graph of small effective diameter. *While the Kronecker power construction in the deterministic case yields graphs with a range of desired properties, its discrete nature produces staircase effects in the degrees and spectral quantities, simply because individual values have large multiplicities [1]*. So authors in [1] propose a stochastic model of Kronecker product and empirically shows it can create *smoother* and more realistic graph than can be generated by its deterministic counter-part. Some basic properties (such as connectivity, existence of giant component, small diameter etc) of stochastic Kronecker graph have been thoroughly investigated in [7]. But we believe the theory of *stochastic Kronecker Graph* is still very young and many obvious questions about it are yet unanswered. Here we try to answer a few question regarding the number of isolated vertices, number of edges, self loops and triangles. In section: 3, we find the expected number of different features as a function of parameters of the stochastic Kronecker graph. Next as obvious corollaries of the above, we establish the sufficient conditions to generate graphs having no isolated vertex, no edges, no self loops and things like these.

## 2. Stochastic Kronecker Graph Model

The Kronecker graph model is defined in its full generality in [1]. But here we concentrate on a specific variant of stochastic Kronecker graph with an initiator matrix of size 2. We adopt the definition provided in [7].

*Definition:* A (stochastic) Kronecker graph is defined by
- An integer k
- A symmetric $2 \times 2$ matrix θ: θ[1,1] = α, θ[1,0] = θ[0,1] = β, θ[0,0] = γ where $0 \leq \gamma \leq \beta \leq \alpha \leq 1$. We call θ the base or initiator matrix.
- The graph has $n = 2^k$ number of vertices where each of the vertices is labeled with a unique bit vector of length $k$. Given two vectors of label ($u_1, u_2, u_3, \ldots, u_k$) and ($v_1, v_2, v_3, \ldots, v_k$) the probability that the edge $(u, v)$ exists is given by: $\prod_i \theta[u_i, v_i]$ independent of the all other edges.
- The weight of a vertex is the number of 1 in its labeling.

The restrictions on the parameters of the base matrix θ has been verified empirically in [1]. If these restrictions are maintained, namely when $\gamma \leq \beta \leq \alpha$, then the resultant Kronecker product does give rise to a (statistically) equivalent real world random network.

## 3. Expected Feature Count

In this section we will find out the expected number of isolated vertices, edges, self loops and triangles. As obvious corollaries to these expected feature count we then establish the sufficient condition to generate large random graphs with no isolated vertex, no edge and no self loops. We start this section with a theorem proved in [7].

**Theorem 1:** *The expected degree of a vertex of weight $l$ is $(\alpha + \beta)^l (\beta + \gamma)^{k-l}$.*

### 3.1. Expected Number of Isolated Vertices

**Theorem 2:** *The expected number of isolated vertices in stochastic Kronecker graph with parameter $\alpha, \beta, \gamma$ is: $< (\frac{2}{e^{\beta+\gamma}})^k$.*

**Proof:** For any vertex $u$ of weight $l$ and any other vertex $v$, let $i$ be the number of bits where $u_b = v_b = 1$ and $j$ be the number of bits where $u_b = 0, v_b = 1$. So there will be $(l - i)$ bit positions where $u_b = 1, v_b = 0$ and $(k - l - j)$ bit positions where $u_b = 0, v_b = 0$. As a result the probability of edge $(u, v)$ to be present is given by: $P[u, v] = \alpha^i \beta^{j+l-i} \gamma^{k-l-j}$. The probability that edge $(u, v)$ is not present is given by: $\overline{P[u,v]} = 1 - \alpha^i \beta^{j+l-i} \gamma^{k-l-j}$. Vertex $v$ is indeed a member of a class of vertices ( let this class be $c_{ij}$ ) and there are $\binom{l}{i} \times \binom{k-l}{j}$ identical vertices [ identical with respect to $u$ ] in this class. So the probability that vertex $u$ is not connected to any vertex of class $c_{ij}$ is given by: $(1 - \alpha^i \beta^{j+l-i} \gamma^{k-l-j})^{(\binom{l}{i} \times \binom{k-l}{j})}$. The value of $i$ varies from $0$ to $l$ and the value of $j$ varies from $0$ to $k - l$. As the edges in stochastic Kronecker graph exist (or not) independently of any other edges, the probability that vertex $u$ is connected to none of the vertices over all possible class $c_{ij}$ is given by:

$\prod_{i=0}^{l} \prod_{j=0}^{k-l} (1 - \alpha^i \beta^{j+l-i} \gamma^{k-l-j})^{(\binom{l}{i} \times \binom{k-l}{j})}$.
$< \prod_{i=0}^{l} \prod_{j=0}^{k-l} e^{-\binom{l}{i} \times \binom{k-l}{j} \alpha^i \beta^{j+l-i} \gamma^{k-l-j}}$. [As $(1 - p)^k < e^{-pk}$]
$= \prod_{i=0}^{l} e^{-\binom{l}{i} \alpha^i \beta^{l-i} \times \sum_{j=0}^{k-l} \binom{k-l}{j} \beta^j \gamma^{k-l-j}}$
$= \prod_{i=0}^{l} e^{-\binom{l}{i} \alpha^i \beta^{l-i} \times (\beta+\gamma)^{k-l}}$
$= e^{-(\alpha+\beta)^l (\beta+\gamma)^{k-l}}$
$< e^{-(\beta+\gamma)^k}$
[As $\gamma \leq \beta \leq \alpha$]

Now we define indicator random variables $X_1, X_2, X_3, \ldots, X_n$ where $X_i$ denotes the event that vertex $i$ be isolated and $X$ be the total number of isolated vertices in a random realization. Then, $X = X_1 + X_2 + X_3 + \ldots + X_n$.
So,
$E[X] = \sum_{i=1}^{n} E[X_i] = \sum_{i=1}^{n} P[X_i] < \sum_{l=0}^{k} \binom{k}{l} e^{-(\beta+\gamma)^k} = 2^k \times e^{-(\beta+\gamma)^k} = (\frac{2}{e^{\beta+\gamma}})^k$. □

**Corollary 1:** *If $\beta + \gamma > \ln(2)$ then the stochastic Kronecker graph will have no isolated vertex with high probability[1].*

### 3.2. Expected Number of Edges

**Theorem 3:** *The expected number of edges in stochastic Kronecker graph is: $\frac{1}{2} \times (\alpha + 2\beta + \gamma)^k$.*

**Proof:** From theorem: 1, the expected degree of a vertex of weight $l$ is given by: $(\alpha + \beta)^l (\beta + \gamma)^{k-l}$. And there are $\binom{k}{l}$ number of vertices of weight $l$. Thus summing it over all possible values of $l$ we will be able to calculate the expected total degree of the resultant graph which equals:

$$\sum_{l=0}^{k} \binom{k}{l} (\alpha + \beta)^l (\beta + \gamma)^{k-l}$$
$$= (\alpha + \beta + \beta + \gamma)^k = (\alpha + 2\beta + \gamma)^k$$

---
[1] With high probability, we mean probability $1 - o(1)$.

Now from the degree sum formula [8] the total (expected) number of edges will be $\frac{1}{2}$ times the total (expected) degree. So the expected number of edges $= \frac{1}{2} \times (\alpha + 2\beta + \gamma)^k$. □

***Corollary 2:*** *If $\alpha + 2\beta + \gamma < 1$ then the stochastic Kronecker graph will have no edges with high probability.*

### 3.3. Expected Number of Self Loops

***Theorem 4:*** *The expected number of self loops in stochastic Kronecker graph is $(\alpha + \gamma)^k$.*

***Proof:*** The probability that a vertex of weight $l$ is connected to itself is $\alpha^l \gamma^{k-l}$. Summing it over all possible values of $l$ we get the total expected number of self loops (as the self loops exist independent of each other we can simply sum up their individual probability): $\sum_{l=0}^{k} \binom{k}{l} \alpha^l \gamma^{k-l} = (\alpha + \gamma)^k$. □

***Corollary 3:*** *If $\alpha + \gamma < 1$ then the stochastic Kronecker graph will have no self loops with high probability.*

### 3.4. Expected Number of Triangles

***Theorem 5:*** *The expected number of triangles in stochastic Kronecker graph is:* $< (\beta + \gamma)^k \alpha^k \left( \alpha \frac{\alpha+\beta}{\beta+\gamma} + \beta + \beta \frac{\alpha+\beta}{\beta+\gamma} + \gamma \right)^k$.

***Proof:*** Let us consider three arbitrary vertices $v_1, v_2, v_3$ of weight $l_1, l_2, l_3$ respectively. We now define four variables $i_1, i_2, j_1, j_2$ as follows: $i_1 =$ number of those bit positions where both $v_1, v_2$ have '1' in their labeling; $i_2 =$ number of those bit positions where both $v_2, v_3$ have '1' in their labeling; $j_1 =$ number of bit positions where $v_1 = 0, v_2 = 1$; $j_2 =$ number of bit positions where $v_2 = 0, v_3 = 1$. Now there will be $(l_1-i_1)$ bit positions where $v_1 = 1, v_2 = 0$ and $(l_2-i_2)$ bit positions where $v_2 = 1, v_3 = 0$; $(k-l_1-j_1)$ bit positions where both $v_1, v_2$ have '0' in their labeling and $(k-l_2-j_2)$ bit positions where both of $v_2$ and $v_3$ have '0' labeling. So the probability that both the edge $(v_1, v_2)$ and $(v_2, v_3)$ be present is given by: $\alpha^{i_1} \beta^{j_1+l_1-i_1} \gamma^{k-l_1-j_1} \times \alpha^{i_2} \beta^{j_2+l_2-i_2} \gamma^{k-l_2-j_2}$. Also we notice that $l_2 = i_1 + j_1$ and $l_3 = i_2 + j_2$. Now summing it over all possible values of $i_1, i_2, j_1, j_2$ we will get the total expected number of two length paths:

$$= \sum_{i_1=0}^{l_1} \sum_{j_1=0}^{k-l_1} \binom{l_1}{i_1} \binom{k-l_1}{j_1} \alpha^{i_1} \beta^{j_1+l_1-i_1} \gamma^{k-l_1-j_1}$$
$$\times \sum_{i_2=0}^{l_2} \sum_{j_2=0}^{k-l_2} \binom{l_2}{i_2} \binom{k-l_2}{j_2} \alpha^{i_2} \beta^{j_2+l_2-i_2} \gamma^{k-l_2-j_2}$$

$$= \sum_{i_1=0}^{l_1} \sum_{j_1=0}^{k-l_1} \binom{l_1}{i_1} \binom{k-l_1}{j_1} \alpha^{i_1} \beta^{j_1+l_1-i_1} \gamma^{k-l_1-j_1}$$
$$\times \sum_{i_2=0}^{l_2} \binom{l_2}{i_2} \alpha^{i_2} \beta^{l_2-i_2} \sum_{j_2=0}^{k-l_2} \binom{k-l_2}{j_2} \beta^{j_2} \gamma^{k-l_2-j_2}$$

$$= \sum_{i_1=0}^{l_1} \sum_{j_1=0}^{k-l_1} \binom{l_1}{i_1} \binom{k-l_1}{j_1} \alpha^{i_1} \beta^{j_1+l_1-i_1} \gamma^{k-l_1-j_1}$$
$$\times (\alpha+\beta)^{l_2} (\beta+\gamma)^{k-l_2}$$

$$= \sum_{i_1=0}^{l_1} \sum_{j_1=0}^{k-l_1} \binom{l_1}{i_1} \binom{k-l_1}{j_1} \alpha^{i_1} \beta^{j_1+l_1-i_1} \gamma^{k-l_1-j_1}$$
$$\times \frac{(\alpha+\beta)^{l_2}}{(\beta+\gamma)^{l_2}} (\beta+\gamma)^k$$

$$= \sum_{i_1=0}^{l_1} \sum_{j_1=0}^{k-l_1} \binom{l_1}{i_1} \binom{k-l_1}{j_1} \alpha^{i_1} \beta^{j_1+l_1-i_1} \gamma^{k-l_1-j_1}$$
$$\times \frac{(\alpha+\beta)^{i_1+j_1}}{(\beta+\gamma)^{i_1+j_1}} (\beta+\gamma)^k$$

$$= (\beta+\gamma)^k \sum_{i_1=0}^{l_1} \binom{l_1}{i_1} \left( \alpha \frac{\alpha+\beta}{\beta+\gamma} \right)^{i_1} \beta^{l_1-i_1}$$
$$\times \sum_{j_1=0}^{k-l_1} \binom{k-l_1}{j_1} \left( \beta \frac{\alpha+\beta}{\beta+\gamma} \right)^{j_1} \gamma^{k-l_1-j_1}$$

$$= (\beta+\gamma)^k \left( \alpha \frac{\alpha+\beta}{\beta+\gamma} + \beta \right)^{l_1} \left( \beta \frac{\alpha+\beta}{\beta+\gamma} + \gamma \right)^{k-l_1}$$

A two length path can be easily extended to a three length cycle by simply connecting the two end vertices by an edge. From the definition of stochastic Kronecker graph, we know that the maximum probability of the existence of an edge is $\leq \alpha^k$. So the expected number of triangles starting at $v_1$ will be:

$$\leq (\beta+\gamma)^k \left(\alpha\frac{\alpha+\beta}{\beta+\gamma}+\beta\right)^{l_1} \left(\beta\frac{\alpha+\beta}{\beta+\gamma}+\gamma\right)^{k-l_1} \times \alpha^k$$

Now we note that the above quantity indeed counts for some *fictitious* triangles of the form $v_1v_1v_1$ or like $v_1v_2v_2$. So the sign '$\leq$' should be replaced by '$<$'. We also note that every triangle is counted twice in the above: one in the clockwise and the other is the counter clock-wise ordering of its vertices. Now incorporating the above facts and summing it over all possible choices of $v_1$, we will be able to get the total expected number of triangles in a random realization of stochastic Kronecker graph model and this quantity would be:

$$< \sum_{l_1=0}^{k} \binom{k}{l_1} (\beta+\gamma)^k \left(\alpha\frac{\alpha+\beta}{\beta+\gamma}+\beta\right)^{l_1} \left(\beta\frac{\alpha+\beta}{\beta+\gamma}+\gamma\right)^{k-l_1} \alpha^k$$

$$= (\beta+\gamma)^k \alpha^k \sum_{l_1=0}^{k} \binom{k}{l_1} \left(\alpha\frac{\alpha+\beta}{\beta+\gamma}+\beta\right)^{l_1} \left(\beta\frac{\alpha+\beta}{\beta+\gamma}+\gamma\right)^{k-l_1}$$

$$= (\beta+\gamma)^k \alpha^k \left(\alpha\frac{\alpha+\beta}{\beta+\gamma}+\beta+\beta\frac{\alpha+\beta}{\beta+\gamma}+\gamma\right)^k \square$$

## 4. Phase Transitions

In this section we show two phase transitions in stochastic Kronecker graph. In proving the existence of phase transitions we need to resort to the second moment argument which simply says that: If X is a random variable, then $P[X=0] \leq \frac{E(X^2)-E(X)^2}{E(X)^2}$, in particular, $P[X=0] \to 0$ when $\frac{E(X^2)}{E(X)^2} \to 1$.

### 4.1. Appearance of Edges

***Theorem 5***: *The appearance of edges in stochastic Kronecker graph exhibits a threshold at $\alpha + 2\beta + \gamma = 1$.*

**Proof**: Let $X$ be the total number of edges in a random realization of a stochastic Kronecker graph and $X_i, 1 \leq i \leq \binom{n}{2}$ be the indicator random variable for the existence of the $i$-th edge. Now from theorem: 3, we know when $\alpha + 2\beta + \gamma < 1$ then $\lim_{n\to\infty} E(X) \to 0$. So when $\alpha + 2\beta + \gamma > 1$ we only need to show that at that instance, $E(X^2) = E(X)^2$ (which will then complete the proof of existence of a phase transition at $\alpha + 2\beta + \gamma = 1$ utilizing the second moment argument). Here we notice that all the $X_i$ are independent according to the definition of Kronecker graph. Hence,

$$E(X)^2 = E\left(X_1 + X_2 + X_3 + \cdots\cdots + X_{\binom{n}{2}}\right)^2$$
$$\sum_{i,j} E(X_i X_j) = \sum_i E(X_i) \sum_j E(X_j) = E(X)^2 \square$$

### 4.2. Appearance of Self Loops

***Theorem 6***: *The appearance of self loops in stochastic Kronecker graph exhibits a threshold at $\alpha + \gamma = 1$.*

**Proof**: Let $X$ be the total number of self loops in a random realization and $X_i, 1 \leq i \leq n$, be the indicator random variable for the existence of the $i$-the loop. Now from theorem: 4, we know that when $\alpha + \gamma < 1$ then $\lim_{n\to\infty} E(X) \to 0$. So to complete the proof of existence of a threshold at $\alpha + \gamma = 1$ we need to show that when $\alpha + \gamma > 1$, then $E(X^2) = E(X)^2$. Now we note that all the self loops exist independent of one another. So we have:

$$E(X)^2 = E(X_1 + X_2 + X_3 + \cdots\cdots + X_n)^2$$
$$= \sum_{i,j} E(X_i X_j) = \sum_i E(X_i) \sum_j E(X_j) = E(X)^2 \square$$

## 5. Conclusion

The stochastic Kronecker model of graph generation is very new and most of the properties of graphs generated by this model are yet to be investigated. Here we try to explore some of its properties namely expected number of isolated vertex, edge, self loop, triangles along with two phase transitions. Based on these expected feature counts, we then establish some of the sufficient conditions to generate graphs with an interesting set of properties.

**Ahmed Mehedi Nizam** has obtained his B.Sc. Engineering degree from Bangladesh University of Engineering and Technology (BUET), Dhaka 1000, on February 2011 and is currently a M.Sc. Engg. Student in the same.

**Md. Nasim Adnan** received his M.Sc. in CSE from Bangladesh University of Engineering and Technology (BUET) and B.Sc. in CSE from Khulna University. Currently he is working as an Adjunct Assistant Professor at the Dept. of Computer Science and Engineering in University of Liberal Arts Bangladesh (ULAB). He also served as a Deputy Director in the Dept. of ITOCD, Bangladesh Bank. His research interest includes Data Mining, Database Systems, Software Engineering, Software and Systems Testing, and E-Commerce.

**Md. Rashedul Islam** received his M.Sc. in IT from RWTH-Achen and B.Sc. in CSE from Khulna University. Currently he is working as a Senior Lecturer at the Dept. of Computer Science and Engineering in University of Liberal Arts Bangladesh (ULAB).

**Mohammod Akbar Kabir** received his M.Sc. and B.Sc. (Hons.) in Computer Science from the University of Dhaka in 2000 and 1998 respectively. Currently he is working as an Assistant Professor in the Dept. of Economics, University of Dhaka. He also served as an Assistant Programmer in the Dept. of ITOCD, Bangladesh Bank and as a Lecturer in the Dept. of Computer Science, Dhaka City College. His research interests are in the area of VLSI Design and E-Commerce.